# Magnetic-field-induced nonlinear transport in HfTe$_5$

Cheng Zhang[1,3,4#*], Jinshan Yang[5#], Zhongbo Yan[6#], Xiang Yuan[7], Yanwen Liu[3], Minhao Zhao[3], Alexey Suslov[8], Jinglei Zhang[9], Li Pi[9], Zhong Wang[10,11], Faxian Xiu[1,2,3,4,12*]

[1] Institute for Nanoelectronic Devices and Quantum Computing, Fudan University, Shanghai 200433, China
[2] Shanghai Qi Zhi Institute, Shanghai, 200232, China
[3] State Key Laboratory of Surface Physics and Department of Physics, Fudan University, Shanghai 200433, China
[4] Zhangjiang Fudan International Innovation Center, Fudan University, Shanghai 201210, China
[5] State Key Laboratory of High Performance Ceramics and Superfine Microstructure, Shanghai Institute of Ceramics, Chinese Academy of Sciences, Shanghai 200050, China
[6] School of Physics, Sun Yat-Sen University, Guangzhou 510275, China
[7] State Key Laboratory of Precision Spectroscopy, East China Normal University, Shanghai, 200062, China
[8] National High Magnetic Field Laboratory, Tallahassee, Florida 32310, USA
[9] Anhui Province Key Laboratory of Condensed Matter Physics at Extreme Conditions, High Magnetic Field Laboratory of the Chinese Academy of Sciences, Hefei 230031, China
[10] Institute for Advanced Study, Tsinghua University, Beijing 100084, China
[11] Collaborative Innovation Center of Quantum Matter, Beijing 100871, China
[12] Shanghai Research Center for Quantum Sciences, Shanghai 201315, China

[#] These authors contributed equally to this work
[*] Correspondence and requests for materials should be addressed to F. X. (E-mail: Faxian@fudan.edu.cn) or C. Z. (E-mail: Zhangcheng@fudan.edu.cn)

**Abstract**
**The interplay of electron correlations and topological phases gives rise to various exotic phenomena including fractionalization, excitonic instability, and axionic excitation. Recently-discovered transition-metal pentatellurides can reach the ultra-quantum limit in low magnetic fields and serve as good candidates for achieving such a combination. Here, we report evidences of density wave and metal-insulator transition in HfTe$_5$ induced by intense magnetic fields. Using the nonlinear transport technique, we detect a distinct nonlinear conduction behavior in the longitudinal resistivity within the *a-c* plane, corresponding to the formation of a density wave induced by magnetic fields. In high fields, the onset of the nonlinear conduction in the Hall resistivity indicates an impurity-pinned magnetic freeze-out as the possible origin of the insulating behavior. These frozen electrons can be gradually re-activated into mobile states above a threshold electric field. These experimental evidences call for further investigations into the underlying mechanism for the bulk quantum Hall effect and field-induced phase transtions in pentatellurides.**

**Introduction**

Topological phases of matter represent a wide range of electronic systems with nontrivial band topology. New quasiparticles, such as Dirac and Weyl fermions, forming by band crossing points in these topological matters, have been discovered in experiments[1–5]. To date, most topological materials discovered so far fall in the scope of a single-particle picture.[6,7] Owing to the vanishing density of states in the vicinity of band crossing points, the Coulomb interaction between these Dirac/Weyl fermions becomes unscreened with a long-range component, which may present distinct correlation effects[8]. One way to enhance the correlation is by using high magnetic fields to compress dilute electrons into highly degenerate states. By considering different interactions, various scenarios were proposed theoretically[9–15] in interacting topological semimetals. Among them, an important direction is to generate axionic dynamics by inducing density wave (DW) states[9], which forms the quasiparticle of axions[16], one of the most promising candidates for the dark matter. The pursuit of these proposals in experiments requests a low Fermi level so that electrons can be condensed into low-index Landau levels within the accessible field range.

Layered transition-metal pentatellurides $ZrTe_5$ and $HfTe_5$ are recently found to be close to an accidental Dirac semimetal phase in the boundary between strong and weak topological insulators, sensitively affected by the lattice constant[17–22]. While the exact value of the band gap is under debate, the bulk states of these systems can be regarded as Dirac fermions with a small mass term[19,23,24]. Angle-dependent quantum oscillations in $ZrTe_5$ revealed a tiny Fermi surface, with a rod-like ellipsoid shape and large anisotropy in the Fermi wave vector and effective mass between the out-of-plane direction (*b* axis) and the in-plane directions (*a* and *c* axes)[20]. Owing to the easy access to the quantum limit within a moderate field, chiral anomaly[18,25,26], log-periodic quantum oscillations[27–29], anomalous Hall/Nernst effect[22,30–32], and quantized plateaus of Hall resistivity[33–35] have been observed in bulk crystals and flakes of $ZrTe_5$ as well as its cousin $HfTe_5$. These observations suggest pentatellurides as good candidates of topological systems for studying the electron correlation effect in magnetic fields.

**Results**

In this study, by combining linear and nonlinear quantum transport, we present evidences of field-induced DW and insulating states in $HfTe_5$, which can be further modulated by an electrical field bias. Single crystals of $HfTe_5$ were produced by the iodine-assisted chemical vapor transport (CVT) as described in Section I of Supplemental Materials. $HfTe_5$ has a carrier density of $2.7 \times 10^{17}$ cm$^{-3}$ at 2 K, one order of magnitude lower than that of $ZrTe_5$ ($2.5 \times 10^{18}$ cm$^{-3}$) grown by the same method[36]. The low carrier density enables an easy access to the quantum limit. We performed transport experiments with the current along the *a* axis and the magnetic field along the *b* axis noted as $\theta = 0°$ configuration. Figure 1a shows the temperature-dependent resistivity with a peak around 72 K in Sample H1. This peak comes from the shift of Fermi energy from the valence band towards the conduction band as the temperature decreases. A similar transition also occurs in CVT-grown $ZrTe_5$ as indicated by the sign change in both Hall and Seebeck coefficients with temperature[37]. Figure 1b shows the longitudinal magnetoresistivity ($\rho_{xx}$, the red curve) and Hall resistivity ($\rho_{xy}$, the blue curve) with clear oscillations at $\theta = 0°$. Apart from small quantum oscillations at low fields, $\rho_{xx}$ experiences a large peak around 2 T when $\theta = 0°$, followed by a dip at 5.4 T. As marked by the Landau level index *n* at resistivity peaks,

HfTe$_5$ enters the quantum limit around 2 T, where only the lowest zeroth Landau levels are occupied. The oscillation peak at $n=1$ reaches a much larger magnitude compared to the others at lower fields. Meanwhile, $\rho_{xy}$ shows a series of plateau-like features as a function of $B$, in coordinate with the oscillations in $\rho_{xx}$. It resembles the bulk quantum Hall (QH) effect recently observed in ZrTe$_5$[33,38,39] and HfTe$_5$[34,35] but with a finite longitudinal resistivity residue. By tracking the angle dependence of quantum oscillations within the $a$-$b$ and $c$-$b$ planes (Fig. S1 a-b), we can extract oscillation frequencies as functions of angles as shown in Fig. 1c ($\theta$ for the $a$-$b$ plane and $\varphi$ for the $c$-$b$ plane as illustrated in the insets of Fig. 1c). It suggests that the Fermi surface of HfTe$_5$ also adopts an ellipsoid shape (Fig. 1d), similar to that of ZrTe$_5$. The fitting of the oscillation frequencies in Fig. 1c yields a strong anisotropy as $k_a$: $k_b$: $k_c$=1:13.6:2.3, with $k_a$, $k_b$, and $k_c$, the lengths of the three semi-axes in the Fermi ellipsoid, being 0.0045, 0.0613, and 0.0102 Å$^{-1}$, respectively.

We further investigate the transport properties of HfTe$_5$ in higher magnetic fields. Figures 2 a and b show $\rho_{xx}$ and $\rho_{xy}$ of another HfTe$_5$ crystal (Sample H2) measured in a resistive magnet up to 31.5 T at different angles. For $\theta=0°$, the magnetoresistivity in the low-field regime well reproduces the results presented in Fig. 1 with a large dip in $\rho_{xx}$ at ~4 T. Subsequently, $\rho_{xx}$ continues increasing and gradually saturates above 21 T. No further oscillations appear since only the lowest Landau level is occupied. Notably, $\rho_{xy}$ presents an anomalous peak-like feature (marked by the black arrows in Fig. 2b) at 11 T, then starts to decrease and finally saturates above ~21 T. As the field is tilted toward the in-plane direction, the magnetoresistivity ratio gets suppressed and the peak position ($B_p$) in $\rho_{xy}$ moves towards higher magnetic fields. In Fig. S1d, we show that $B_p$ can be fitted by the cosine function of the tilting angle $\theta$, suggesting that the peak of $\rho_{xy}$ is likely to be induced by the out-of-plane component of the magnetic field. Meanwhile, the plateau at $\rho_{xy}=\rho_1$ persists at large $\theta$ as marked by the grey dashed line in Fig. 2b. The temperature dependences of $\rho_{xx}$ and $\rho_{xy}$ are plotted in Figs. 2 c and d, respectively. The suppression of $\rho_{xy}$ in the high-field regime quickly disappears above 15 K and $\rho_{xx}$ decreases drastically, while the low-field parts of $\rho_{xy}$ overlap well at different temperatures. The field dependence of $\rho_{xx}$ at 1.5~6 K shows a crossing point around 11.8 T (Fig. S6). It corresponds to a field-induced metal-insulator transition, which follows a scaling relation with $T$ and $B$ in the vicinity of the crossing point as shown in Section III of Supplementary Materials. The strong temperature dependence (1.5-15 K) of the high-field parts in $\rho_{xx}$ and $\rho_{xy}$ suggests that the peak in $\rho_{xy}$ is unlikely to originate from the trivial multi-carrier transport mechanism. By considering the field dependence of both $\rho_{xx}$ and $\rho_{xy}$, we conclude that HfTe$_5$ enters an insulating state around 11.8 T, where $\rho_{xx}$ is enhanced and $\rho_{xy}$ is suppressed due to the decrease in the number of mobile carriers.

Strong nonlinear transport of HfTe$_5$ is observed in magnetic fields. We started by applying a series of direct currents (DC) (0~1 mA) with a superimposed alternating current (AC) of 50 μA to the sample and detected the AC resistance using the standard lock-in technique. As presented in Figs. 2 e and f, the curves overlap for $I_{DC}=0$ and 0.25 mA. As $I_{DC}$ exceeds 0.5 mA, the suppression of $\rho_{xy}$ at high fields is gradually eliminated and $\rho_{xx}$ decreases as well in the same regime. The bias-dependent resistivity indicates that the insulating phase of HfTe$_5$ at high fields is suppressed upon large biases.

To track the evolution of the nonlinear behavior, we investigated the bias-dependent DC differential resistivity. In Figs. 3 a and b, the differential resistivity ($\rho_{xx}$ and $\rho_{xy}$) is plotted as a function of the biased electric field $E_b$ above -15 T. Under small biases, the system remains in the linear transport regime with differential resistivity values fixed at different biases. Sharp transitions from linear to nonlinear regimes occur above a threshold electric field, $E_T$. In the range of -15~-31 T, $\rho_{xx}$ ($\rho_{xy}$) shows a prominent dip (peak) at $E_T$ followed by a linear decrease (increase) upon the increase of $E_b$. In contrast, in the range of -5~-10 T, $\rho_{xx}$ increases with $E_b$ beyond the threshold, while $\rho_{xy}$ becomes nearly independent of $E_b$ except for a slight variation near $E_T$. It is in stark contrast to the conventional joule heating effect induced by large currents. The suppression of in-plane transport is gradually relieved by increasing $E_b$ and the carriers become mobile again. The sharp transition near the onset of the nonlinearity will smear out as the temperature increases but the nonlinear transport persists (Fig. S5c). One unexpected finding is that $\rho_{xx}$ continues to show nonlinear transport in the range of 0~-10 T (where the linear transport shows no anomaly) while $\rho_{xy}$ does not (Figs. 3 c-d and Fig. S5a). At zero magnetic field, the nonlinear behavior vanishes (*i.e.*, the resistivity becomes independent of the applied biases).

In Figs. 4 a-b, we extract the threshold electric field $E_T$ and the relative resistivity change $\Delta\rho/\rho$ near the transition as a function of magnetic fields. The relative resistivity change quantifies the effect on the transport property. The difference in nonlinear conduction of diagonal and off-diagonal components in resistivity tensors separates the phase diagram of HfTe$_5$ under magnetic fields into two regimes. In Regime I, HfTe$_5$ remains metallic and only diagonal resistivity from longitudinal transport is affected by bias fields (down-left inset in Fig. 4b). This behavior is consistent with the widely studied sliding motion of DW states. A bias exceeding $E_T$ results in a one-dimensional axial motion and does not contribute to Hall effect[40,41]. The differential resistivity curves shown in Fig. 3 are symmetric with electric fields and change linearly after the depinning transition, while other origins of nonlinear resistivity such as hopping conduction or *p-n* junctions give dramatic different behaviors (activation behavior in electric fields for hopping conduction[42], nonsymmetric I-V curve for *p-n* junctions or other interface potential barriers[43]). The sharp dip/peak features near the transition and the threshold electric field value are consistent with the typical sliding motion of density wave systems.[44,45,45,46] In Regime II, electrons in HfTe$_5$ becomes strongly localized and a large bias can activate both diagonal and off-diagonal resistivity tensors, making the system metallic again. Such behavior is distinct from the sliding DW case and fits the depinning process of defect-localized electron solid states. A large bias breaks the binding between electrons and impurities and makes electrons mobile again. Then both longitudinal and Hall transport ($\rho_{xx}$ and $\rho_{xy}$) will be activated (top-right inset in Fig. 4b). We note that although a classical picture involving the "backflow" of normal electrons can affect Hall resistivity in the sliding DW state[47], it leads to a decrease rather than an increase of $|\rho_{xy}|$ and cannot give such a large change in $\Delta\rho_{xy}/\rho_{xy}$, contradictory to our observation.

**Discussion**

In a previous study[33], the emergence of bulk QH effect in ZrTe$_5$ was interpreted as DW forming along the *b* direction induced by Landau level nesting. Similar to ZrTe$_5$, HfTe$_5$ shows signatures of QH effect[34,35], which indicates that the interlayer dispersion is diminished. Nevertheless, the Landau level nesting picture cannot account for the in-plane nonlinear transport since it only induces DW along the *b* direction. The absence of the sliding behavior at zero field and the systematic increase of $E_T$ with *B* indicate that the DW state is induced by magnetic fields. The picture of field-induced spin DW as observed in organic conductors[48] matches our results in a better way, while the charge DW order will normally be suppressed by the magnetic fields[49]. According to the quantized nesting model[48], the strength of DW is governed by the Fermi surface anisotropy (refer to Section IV of Supplementary Materials for detailed discussions). Therefore, the interlayer periodic potential confines the system into a series of 2D planes, leading to the emergence of the bulk QH effect. Above 2 T, the carriers are confined to the zeroth Landau levels. The in-plane transport can be evaluated via the bias-dependent differential resistivity measurements. Only a small part of carriers is localized by the DW at low fields as evidenced by the relatively small change of the resistivity between the pinning and sliding states. This conclusion also explains another unusual behavior – the sign of $\Delta\rho_{xx}/\rho_{xx}$ oscillating with quantum oscillations at low fields shown in Fig. S5a. Both the normal and DW electrons coexist in the system with the former being the majority. Under large biases beyond $E_T$, the DW electrons start one-dimensionally sliding motion and contribute to the conduction. Then according to the two-fluid model, the current of normal electrons becomes smaller since the total current is fixed. Therefore, the quantum oscillations given by normal electron conduction will become comparably weaker, which results in the oscillating sign of $\Delta\rho_{xx}/\rho_{xx}$ with $B$ as shown in Fig. S5a.

Based on these observations, we propose a possible scenario for the phase diagram of $HfTe_5$ in magnetic fields. The gap size of $HfTe_5$ decreases with temperature, resulting in the low-temperature phase close to an accidental Dirac semimetal.[19,21] At zero field, the Fermi surface of $HfTe_5$ resembles a highly anisotropic ellipsoid filled with electrons. With magnetic fields along the $b$ direction, the system is firstly driven to a three-dimensional density wave state, then gradually collapses into an impurity-pinned insulating state around 11.8 T. Generally, there are mainly two kinds of mechanisms accounting for the defect-pinned insulating phase in the quantum limit of dilute electronic systems. One is the formation of Wigner crystals[50]. It is a collective state where the potential energy of electrons dominates the kinetic energy and the electrons crystallize into a lattice. However, the typical temperature required to form a Wigner crystal is in the milli-Kelvin range while the insulating phase in $HfTe_5$ persists up to over 15 K. Besides, the critical field of Wigner transition was found to be temperature dependent[51] while the transition features in $HfTe_5$ were found to be temperature insensitive. The other one is the magnetic freeze-out effect, which originates from the electron-impurity interaction[52]. In this case, electrons bind to impurities due to the small magnetic length at high fields (Fig. 4c). In Section IV of Supplementary Materials, a rough estimation based on the effective mass and the carrier density in $HfTe_5$ gives the value field of the magnetic freeze-out $B_c \approx 8.2$ T, close to the experimental results. In the view of band structure, the Fermi level meets the impurity band at high magnetic fields and electrons are localized near the impurity sites as illustrated in Fig. 4d. Thus, the Hall conductivity should be thermally activated by temperature as $\sigma_{xy} \propto e^{-\varepsilon_b/kT}$ with $\varepsilon_b$ the field-dependent binding energy and $k$ the Boltzmann constant. As shown in Fig. S6, the absolute value of the Hall conductivity increases with the temperature above 14 T. Figure 4e shows the activation of $\sigma_{xy}$ in different fields. The fitted activation gap $\Delta$ as a function of the square root of the magnetic field is plotted in the inset of Fig. 4e. It increases linearly with $\sqrt{B}$ in the range of 14.5~20 T and gradually saturates to ~0.4 meV after that, in agreement with the magnetic freeze-out case as detailed discussed in Section IV of Supplementary Materials.

One interesting feature is that $\rho_{xx}$ only changes by less than 40% at different biases while $\rho_{xy}$ is suppressed to nearly zero in the high-field localized states. In Fig. S7, we show that in high fields the Hall component of conductivity gradually exhibits different temperature dependence from the longitudinal part. While further investigations are on demand, it may be related to possible 1D edge modes as proposed recently in KHgSb[53]. The observations of in-plane nonlinear conduction and possible magnetic freeze-out state indicate a complicated phase diagram of HfTe$_5$ in magnetic fields. Thermodynamic properties and the *ac* transport of sliding DW may be investigated to further identify the nature of these phase transitions. And experimental methods sensitive to spin-polarization can be applied to settle down whether it is a charge or spin DW. Meanwhile, the unique electromagnetic response (axion electrodynamics) in topological systems allows for the fluctuation of topological *θ*-term in the presence of the DW order, which behaves like the axion and gives rise to novel topologically protected transport properties.[9,54] The DW state of HfTe$_5$ in magnetic fields can be therefore used to explore the physics of axionic dynamics[9,12,54].

**Methods**

High-quality single crystals of HfTe$_5$ were grown via chemical vapor transport with iodine as the transport agent, similar to ZrTe$_5$[20]. Stoichiometric Hafnium flakes (99.98%, Aladdin) and Tellurium powder (99.999%, Alfa Aesar) were ground together and sealed in an evacuated quartz tube with iodine flakes (99.995%, Alfa Aesar). A temperature gradient of 60 °C between 490 and 430 °C in a two-zone furnace was used for crystal growth. The typical as-grown sample has a long ribbon-like shape along *a*-axis. The sample crystalline quality and stoichiometry were checked by the X-ray diffraction and the energy dispersive spectrum. The magneto-transport measurements were performed in a Physical Property Measurement System (Quantum Design) for the low fields and in resistive magnets in Tallahassee, US and Hefei, China for the high fields with the standard lock-in technique. The differential resistivity measurements were carried out using the corresponding in-build mode of Keithley 6221 and 2182 models. The bias electric field is extracted as $E_b = IR/L$ with *I*, *R* and *L* being the applied current, four-terminal resistance value, and the channel length between two inner voltage-sensing electrodes, respectively.

**Acknowledgements**
F.X. was supported by the National Natural Science Foundation of China (11934005 and 11874116), National Key Research and Development Program of China (Grant No. 2017YFA0303302 and 2018YFA0305601), the Science and Technology Commission of Shanghai (Grant No. 19511120500), the Shanghai Municipal Science and Technology Major Project (Grant No. 2019SHZDZX01), and the Program of Shanghai Academic/Technology Research Leader (Grant No. 20XD1400200). C.Z. was sponsored by the National Natural Science Foundation of China (Grant No. 12174069), Shanghai Sailing Program (Grant No. 20YF1402300), Natural Science Foundation of Shanghai (Grant No. 20ZR1407500) and the start-up grant at Fudan University. Part of the sample fabrication was performed at Fudan Nano-fabrication Laboratory. Part of transport measurements was performed at the High Magnetic Field Laboratory, CAS. A portion of this work was performed at the National High Magnetic Field Laboratory (USA), which is supported by National Science Foundation Cooperative Agreement No. DMR-1644779, No. DMR-1157490 and the State of Florida. Z. Y. was supported by the National Natural Science Foundation of China (Grant No. 11904417), and Natural Science Foundation of Guangdong Province (Grant No. 2021B1515020026). Z.W. was supported by National Natural Science Foundation of China (Grant No. 11674189). J.Z. was supported by Scientific Instrument Developing Project of the Chinese Academy of Sciences (Grant No. YJKYYQ20180059), the Natural Science Foundation of China (Grants No. U1932154), and Youth Innovation Promotion Association CAS (Grant No. 2018486). C.Z. thanks Sihang Liang for helpful discussion.

## Author contributions
F.X. and C.Z. conceived the ideas and supervised the overall research. Y.L., X.Y. and C.Z. synthetized HfTe$_5$ crystals. C.Z., X.Y., Y.L. and M.Z. carried out the transport measurements. A.S., J.Z. and L.P. helped with the transport experiments at high magnetic fields. C.Z. analyzed the transport data. Y.Z. and Z.W. provided theoretical support. C.Z., J. Y. and F.X. wrote the paper with help from all other co-authors.

## Competing financial interests
The authors declare no competing financial interests.

## References

1 Wan X, Turner AM, Vishwanath A, Savrasov SY. Topological semimetal and Fermi-arc surface states in the electronic structure of pyrochlore iridates. *Phys Rev B* 2011; **83**: 205101.
2 Weng H, Fang C, Fang Z, Bernevig BA, Dai X. Weyl Semimetal Phase in Noncentrosymmetric Transition-Metal Monophosphides. *Phys Rev X* 2015; **5**: 011029.
3 Liu ZK, Zhou B, Zhang Y, Wang ZJ, Weng HM, Prabhakaran D *et al.* Discovery of a Three-Dimensional Topological Dirac Semimetal, Na3Bi. *Science* 2014; **343**: 864.
4 Xu S-Y, Belopolski I, Alidoust N, Neupane M, Bian G, Zhang C *et al.* Discovery of a Weyl fermion semimetal and topological Fermi arcs. *Science* 2015; **349**: 613.
5 Lv BQ, Weng HM, Fu BB, Wang XP, Miao H, Ma J *et al.* Experimental Discovery of Weyl Semimetal TaAs. *Phys Rev X* 2015; **5**: 031013.
6 Ando Y. Topological Insulator Materials. *J Phys Soc Jpn* 2013; **82**: 102001.
7 Armitage NP, Mele EJ, Vishwanath A. Weyl and Dirac semimetals in three-dimensional solids. *Rev Mod Phys* 2018; **90**: 015001.
8 Rachel S. Interacting topological insulators: a review. *Rep Prog Phys* 2018; **81**: 116501.
9 Wang Z, Zhang S-C. Chiral anomaly, charge density waves, and axion strings from Weyl semimetals. *Phys Rev B* 2013; **87**: 161107.
10 Sun X-Q, Zhang S-C, Wang Z. Helical Spin Order from Topological Dirac and Weyl Semimetals. *Phys Rev Lett* 2015; **115**: 076802.
11 Wei H, Chao S-P, Aji V. Excitonic Phases from Weyl Semimetals. *Phys Rev Lett* 2012; **109**: 196403.
12 Roy B, Sau JD. Magnetic catalysis and axionic charge density wave in Weyl semimetals. *Phys Rev B* 2015; **92**: 125141.
13 Kim P, Ryoo JH, Park C-H. Breakdown of the Chiral Anomaly in Weyl Semimetals in a Strong Magnetic Field. *Phys Rev Lett* 2017; **119**: 266401.
14 Witczak-Krempa W, Knap M, Abanin D. Interacting Weyl Semimetals: Characterization via the Topological Hamiltonian and its Breakdown. *Phys Rev Lett* 2014; **113**: 136402.
15 Pan Z, Shindou R. Ground-state atlas of a three-dimensional semimetal in the quantum limit. *Phys Rev B* 2019; **100**: 165124.
16 Feng JL. Dark matter candidates from particle physics and methods of detection. *Annu Rev Astron Astrophys* 2010; **48**: 495–545.
17 Weng H, Dai X, Fang Z. Transition-Metal Pentatelluride ZrTe5 and HfTe5: A Paradigm for Large-Gap Quantum Spin Hall Insulators. *Phys Rev X* 2014; **4**: 011002.
18 Li Q, Kharzeev DE, Zhang C, Huang Y, Pletikosić I, Fedorov AV *et al.* Chiral magnetic effect in ZrTe5. *Nat Phys* 2016; **12**: 550.
19 Zhang Y, Wang C, Liu G, Liang A, Zhao L, Huang J *et al.* Temperature-induced Lifshitz transition in topological insulator candidate HfTe5. *Sci Bull* 2017; **62**: 950–956.
20 Yuan X, Zhang C, Liu Y, Narayan A, Song C, Shen S *et al.* Observation of quasi-two-dimensional Dirac fermions in ZrTe5. *NPG Asia Mater* 2016; **8**: e325.
21 Fan Z, Liang Q-F, Chen YB, Yao S-H, Zhou J. Transition between strong and weak topological insulator in ZrTe5 and HfTe5. *Sci Rep* 2017; **7**: 45667.
22 Liang T, Lin J, Gibson Q, Kushwaha S, Liu M, Wang W *et al.* Anomalous Hall effect in ZrTe5. *Nat Phys* 2018; **14**: 451–455.
23 Chen Z-G, Chen RY, Zhong RD, Schneeloch J, Zhang C, Huang Y *et al.* Spectroscopic evidence for bulk-band inversion and three-dimensional massive Dirac fermions in ZrTe5. *Proc Natl Acad*

*Sci* 2017; **114**: 816–821.
24 Jiang Y, Dun ZL, Zhou HD, Lu Z, Chen K-W, Moon S *et al.* Landau-level spectroscopy of massive Dirac fermions in single-crystalline ZrTe5 thin flakes. *Phys Rev B* 2017; **96**: 041101.
25 Wang H, Li C-K, Liu H, Yan J, Wang J, Liu J *et al.* Chiral anomaly and ultrahigh mobility in crystalline HfTe5. *Phys Rev B* 2016; **93**: 165127.
26 Zhao L-X, Huang X-C, Long Y-J, Chen D, Liang H, Yang Z-H *et al.* Anomalous Magneto-Transport Behavior in Transition Metal Pentatelluride HfTe5. *Chin Phys Lett* 2017; **34**: 037102.
27 Wang H, Liu H, Li Y, Liu Y, Wang J, Liu J *et al.* Discovery of log-periodic oscillations in ultraquantum topological materials. *Sci Adv* 2018; **4**: eaau5096.
28 Wang H, Liu Y, Liu Y, Xi C, Wang J, Liu J *et al.* Log-periodic quantum magneto-oscillations and discrete-scale invariance in topological material HfTe5. *Natl Sci Rev* 2019; **6**: 914–920.
29 Liu Y, Wang H, Zhu H, Li Y, Ge J, Wang J *et al.* Tunable discrete scale invariance in transition-metal pentatelluride flakes. *Npj Quantum Mater* 2020; **5**: 88.
30 Zhang JL, Wang CM, Guo CY, Zhu XD, Zhang Y, Yang JY *et al.* Anomalous Thermoelectric Effects of ZrTe5 in and beyond the Quantum Limit. *Phys Rev Lett* 2019; **123**: 196602.
31 Sun Z, Cao Z, Cui J, Zhu C, Ma D, Wang H *et al.* Large Zeeman splitting induced anomalous Hall effect in ZrTe5. *Npj Quantum Mater* 2020; **5**: 36.
32 Liu Y, Wang H, Fu H, Ge J, Li Y, Xi C *et al.* Induced anomalous Hall effect of massive Dirac fermions in ZrTe5 and HfTe5 thin flakes. *Phys Rev B* 2021; **103**: L201110.
33 Tang F, Ren Y, Wang P, Zhong R, Schneeloch J, Yang SA *et al.* Three-dimensional quantum Hall effect and metal-insulator transition in ZrTe5. *Nature* 2019; **569**: 537–541.
34 Wang P, Ren Y, Tang F, Wang P, Hou T, Zeng H *et al.* Approaching three-dimensional quantum Hall effect in bulk HfTe5. *Phys Rev B* 2020; **101**: 161201.
35 Galeski S, Zhao X, Wawrzyńczak R, Meng T, Förster T, Lozano PM *et al.* Unconventional Hall response in the quantum limit of HfTe5. *Nat Commun* 2020; **11**: 5926.
36 Liu Y, Yuan X, Zhang C, Jin Z, Narayan A, Luo C *et al.* Zeeman splitting and dynamical mass generation in Dirac semimetal ZrTe5. *Nat Commun* 2016; **7**: 12516.
37 Shahi P, Singh DJ, Sun JP, Zhao LX, Chen GF, Lv YY *et al.* Bipolar Conduction as the Possible Origin of the Electronic Transition in Pentatellurides: Metallic vs Semiconducting Behavior. *Phys Rev X* 2018; **8**: 021055.
38 Wang W, Zhang X, Xu H, Zhao Y, Zou W, He L *et al.* Evidence for Layered Quantized Transport in Dirac Semimetal ZrTe5. *Sci Rep* 2018; **8**: 5125.
39 Galeski S, Ehmcke T, Wawrzynczak R, Lozano P, Brando M, Kuchler R *et al.* Origin of the quasi-quantized Hall effect in ZrTe5. *Nat Commun* 2020; **12**: 3197.
40 Monçeau, P. P, Ong NP, Portis AM, Meerschaut A, Rouxel J. Electric Field Breakdown of Charge-Density-Wave-Induced Anomalies in NbSe3. *Phys Rev Lett* 1976; **37**: 602–606.
41 Tessema GX, Ong NP. Hall effect of sliding condensate in NbSe3. *Phys Rev B* 1981; **23**: 5607–5610.
42 Yu D, Wang C, Wehrenberg BL, Guyot-Sionnest P. Variable Range Hopping Conduction in Semiconductor Nanocrystal Solids. *Phys Rev Lett* 2004; **92**: 216802.
43 Neamen DA. *Semiconductor physics and devices*. McGraw-Hill higher education, 2003.
44 Thorne RE, Tucker JR, Bardeen J. Experiment versus the classical model of deformable charge-density waves: Interference phenomena and mode locking. *Phys Rev Lett* 1987; **58**: 828–831.
45 Grüner G. The dynamics of charge-density waves. *Rev Mod Phys* 1988; **60**: 1129.
46 Monceau P. Electronic crystals: an experimental overview. *Adv Phys* 2012; **61**: 325–581.
47 Forró L, Cooper JR, Jánossy A, Kamarás K. Nonlinear Hall effect in K0.3MoO3 due to the sliding of charge-density waves. *Phys Rev B* 1986; **34**: 9047–9050.
48 Chaikin PM. Field Induced Spin Density Waves. *J Phys I* 1996; **6**: 1875–1898.
49 Grüner G. The dynamics of spin-density waves. *Rev Mod Phys* 1994; **66**: 1–24.
50 Andrei EY, Deville G, Glattli DC, Williams FIB, Paris E, Etienne B. Observation of a Magnetically Induced Wigner Solid. *Phys Rev Lett* 1988; **60**: 2765–2768.
51 Rosenbaum TF, Field SB, Nelson DA, Littlewood PB. Magnetic-Field-Induced Localization Transition in HgCdTe. *Phys Rev Lett* 1985; **54**: 241–244.

52  Dyakonov MI, Efros AL, Mitchell DL. Magnetic Freeze-Out of Electrons in Extrinsic Semiconductors. *Phys Rev* 1969; **180**: 813–818.
53  Liang S, Kushwaha S, Gao T, Hirschberger M, Li J, Wang Z *et al.* A gap-protected zero-Hall effect state in the quantum limit of the non-symmorphic metal KHgSb. *Nat Mater* 2019; **18**: 443–447.
54  Gooth J, Bradlyn B, Honnali S, Schindler C, Kumar N, Noky J *et al.* Axionic charge-density wave in the Weyl semimetal (TaSe4)2I. *Nature* 2019; **575**: 315–319.

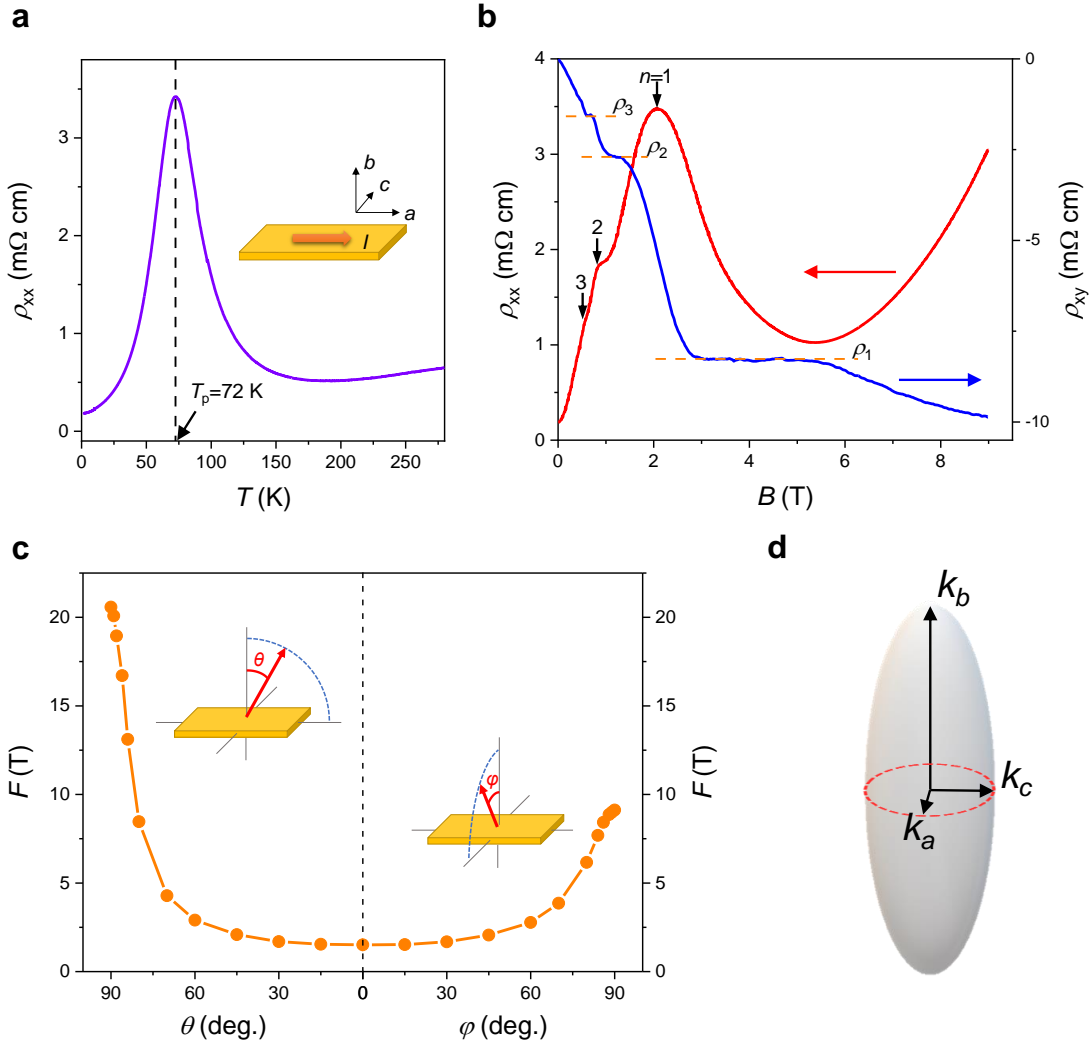

**Figure 1. Magneto-transport and Fermi surface anisotropy of HfTe$_5$.** (**a**) The zero-field resistivity of HfTe$_5$ as a function of $T$ with a peak at 72 K. The current is applied along the $a$ direction within the $a$-$c$ plane. (**b**) The longitudinal ($\rho_{xx}$, the red curve) and Hall ($\rho_{xy}$, the blue curve) resistivity at 2 K with $n$ the Landau level index. Note that the original longitudinal and Hall resistance data of this figure (without normalization with sample geometry) has been presented in Fig. S18d of Ref. [36] by part of the authors. The dashed lines mark the position of Hall plateaus. (**c**) The oscillation frequencies as a function of $\theta$ and $\varphi$. The insets are the measurement geometries. (**d**) The sketch of the Fermi surface in HfTe$_5$.

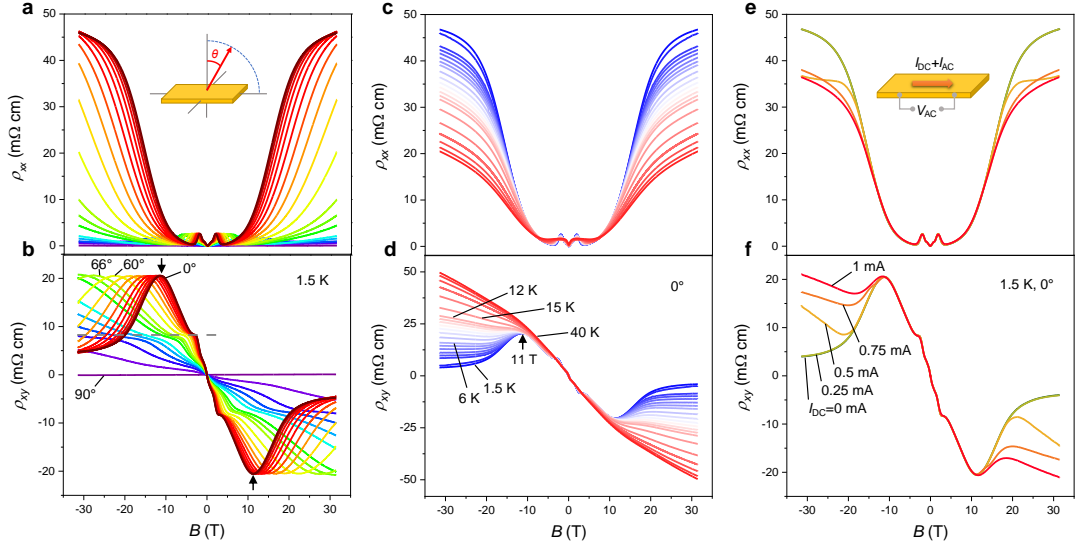

**Figure 2. Linear and nonlinear transport results of HfTe$_5$ at high magnetic fields.** (**a-b**) $\rho_{xx}$ (**a**) and $\rho_{xy}$ (**b**) as a function of $B$ at different angles at 1.5 K. The inset of **a** is the measurement geometry. The intervals between curves in the ranges $\theta=0°\sim 60°$ and $\theta=66°\sim90°$ are 6° and 3°, respectively. The transition in Hall effect is marked by the black arrows. (**c-d**) $\rho_{xx}$ (**c**) and $\rho_{xy}$ (**d**) at different temperatures at $\theta=0°$. The intervals between curves in the ranges 1.5~6 K, 6~12 K and 15~40 K are 0.5 K, 1 K and 5 K, respectively. The crossing point of $\rho_{xx}$ and the transition in $\rho_{xy}$ are marked by the black arrows. (**e-f**) $\rho_{xx}$ (**e**) and $\rho_{xy}$ (**f**) at different DC currents at $\theta=0°$ and 1.5 K. The inset of **e** is the measurement configuration. Each color represents the same $\theta$, temperature, current for **a** and **b**, **c** and **d**, **e** and **f**, respectively. Note that the field-dependent magnetoresistivity data in this figure have been symmetrized (or anti- symmetrized for Hall effect) in $B$ with the original data shown in Fig. S4.

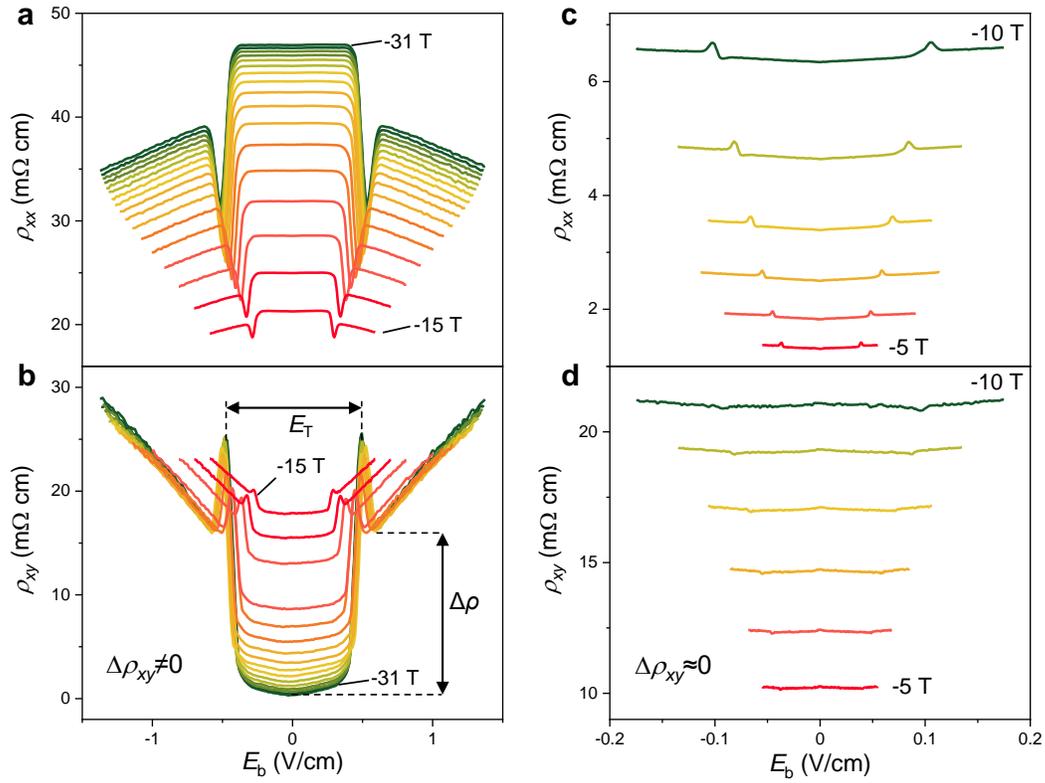

**Figure 3. Bias-dependence of resistivity tensors at different magnetic fields. (a-b)** $\rho_{xx}$ (**a**) and $\rho_{xy}$ (**b**) as a function of $E_b$ fields at 1.5 K. The interval between curves is 1 T. (**c-d**) $\rho_{xx}$ (**c**) and $\rho_{xy}$ (**d**) as a function of $E_b$ at 1.5 K. The interval between curves is 1 T. The value of $E_b$=1 V/cm corresponds to a DC current of 1.09 mA at 31 T.

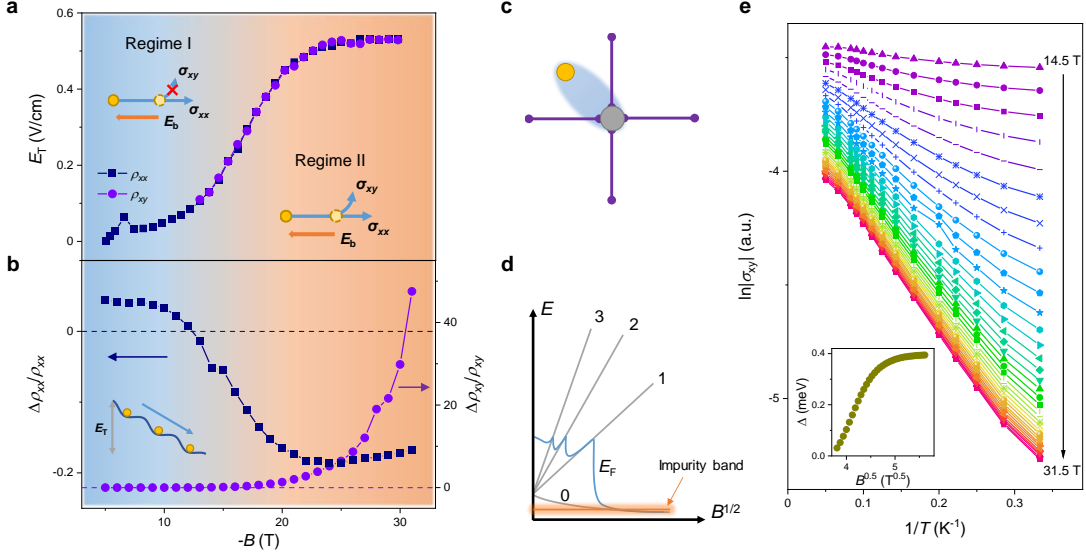

**Figure 4. The DW and insulating phases of HfTe$_5$ in magnetic fields.** (**a-b**) $E_T$ (**a**) and $\Delta\rho/\rho$ (**b**) as a function of $B$. The navy blue and purple dots correspond to $\rho_{xx}$ and $\rho_{xy}$, respectively. $\Delta\rho/\rho$ is the resistivity difference $\Delta\rho$ before and after the transition (Fig. 3b) divided by the zero-$E_b$ value of the resistivity. The insets in **a** show the nonlinear behavior in $\rho_{xy}$, which separates the phase diagram of HfTe$_5$ into two regimes. In Regime I, only the longitudinal transport will be activated. In Regime II, both longitudinal and Hall transport will be activated when the electric field breaks the binding between electrons and impurities. The inset in **b** is a schematic of the DW electron-sliding. (**c**) The magnetic freeze-out state where the carriers bind to the local impurities (the grey circle). (**d**) The schematic band structure for the magnetic freeze-out effect. At high fields the itinerant electrons are pinned to the impurity band. (**e**) Arrhenius plots of $|\sigma_{xy}|$. The inset is the activation gap as a function of $B^{0.5}$.